\documentclass[12pt,a4paper]{article}

\usepackage{graphicx,here}

\def\Ba{\begin{eqnarray}}
\def\Ea{\end{eqnarray}}
\def\Be{\begin{equation}}
\def\Ee{\end{equation}}

\begin{document}

\begin{center}

{\Large Domain Wall Renormalization Group Analysis \\ of 2-dimensional Ising Model}

\vspace{5mm}

{\small Ken-Ichi Aoki \footnote{aoki@hep.s.kanazawa-u.ac.jp}, Tamao Kobayashi \footnote{ballblue@hep.s.kanazawa-u.ac.jp}
and Hiroshi Tomita \footnote{t\_hirosi@hep.s.kanazawa-u.ac.jp}}

\vspace{5mm}

{\small \it Institute for Theoretical Physics, Kanazawa University,\\
Kamuma, Kanazawa, 920-1192, Japan}

\vspace{5mm}

{\bf \small abstract}
\end{center}

{\small Using a recently proposed new renormalization group method (tensor
renormalization group), 
we analyze the Ising model on the 2-dimensional square lattice.
For the lowest order approximation with two domain wall states, 
it realizes the idea of coarse graining of domain walls.
We write down explicit analytic renormalization transformation and
prove that the picture of the coarse graining of the  physical domain walls
does hold for all physical renormalization group flows.
We solve it to get the fixed point structure and obtain the critical exponents and 
the critical temperature. These results are very near to the exact values.
We also briefly report the improvement using four domain wall states.}

\vspace{10mm}

\section{Introduction}

Recently a new type of renormalization group method called 
the tensor renormalization group (TRG) is introduced by Levin and Nave
\cite{TRG} and has been applied to various models, particularly
classical and quantum spin models in 
2-dimension\cite{TRG-triangle,Tensor-entanglement-square-transverse-ising}.
In this article we concentrate on the square lattice Ising model.
The tensor renormalization group method for spin systems 
can be interpreted as the domain wall renormalization group.
We hope it might give deep insight for the idea of the coarse graining
of such topological objects like conserved walls.

Our aim here is to clarify in detail the properties of this domain wall 
renormalization group transformation in the lowest order approximation
with only two domain wall states on the coarse grained links. 
We will be able to write down the explicit analytic form of 
the renormalization group transformation (RGT) 
with the help of the physical region condition.
This analytic formula will help us to prove that the picture of the corase grained 
domain walls does hold for all renormalization group flows starting
from the physical points. 

Then we analytically prove the existence of a single non-trivial fixed point 
controlling the phase transition and evaluate the correlation length
critical exponent which is actually unexpectedly good value, that is, 
within 2\% error of the exact value.

We also obtain the critical temperature within 10\% error of the 
exact value. Using this lowest order RGT 
we also numerically calculate the partition function for finite size systems
as a function of the temperature.
In the infinite size limit, we observe the logarithmic divergence of the 
specific heat in the neighborhood of the criticality.

We finally mention quickly that enlarging the coarse grained state space we set up
4-state version of the domain wall renormalization group and it gives an yet better
result for the critical temperature within 2\% error.

We would like to stress here about new arguments and results in our study.
We obtain explicit analytic forms of the renormalization group transformation and 
solve its fixed point condition to have a single non-trivial fixed point solution.
Although we work in the lowest order approximation, this is a good example of
partially solvable renormalization group equation to clarify total phase structure.
Analytic forms give us much greater inspiration than numerical calculations in the
black box, and we hope it will contribute to physical formulation of 
renormalization group equation for topological objects in the higher order
approximation.
In this course of calculation it is important that we define 
the physical region condition of
parameters, because the condition assures that the notion of the conserved domain walls
hold also for coarse grained variables. Due to this fact, the coarse grained
variables can be regarded as to describe the coarse grained domain wall configurations, 
and therefore we can call the renormalization 
group as the domain wall renormalization group.
Also the critical behaviors we evaluated for this system of the square lattice Ising model
has not been reported explicitly elsewhere.

\section{Domain wall representation and RGT}

\begin{figure}
\begin{center}
\includegraphics[bb=0 0 517 133, width=9cm, clip]{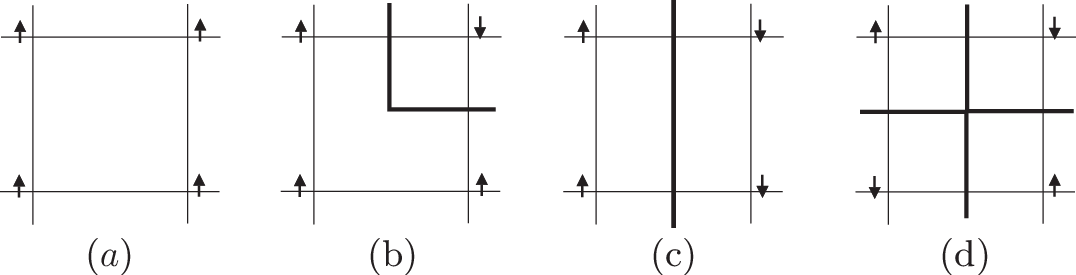}
\nobreak
\caption{Domain wall representation of the square lattice Ising model}
\label{fig:domainwallrep}
\end{center}
\end{figure}

It is well-known that in the 2-dimensional Ising model, any spin configuration
is  equivalently represented by a domain wall configuration defined on the
dual link of the dual lattice. 
As for the square lattice, the dual lattice is also square.
Domain walls are said to exist on a dual link when two spins separated by 
the dual link are different (Fig.\ref{fig:domainwallrep}).

\noindent
Note that a domain wall configuration represents two spin configurations 
of Z$_2$ pair.
Thus summing up all domain wall configuration gives just a half of the
partition function and this factor 1/2 does not affect physical 
quantities discussed below.

We define the partition function of the Ising spin system by
\Be
Z = \sum_{\sigma} \exp\left(\beta J \sum_{\mbox{\tiny n.n.}}
\sigma_i\sigma_j \right)\ .
\Ee
According to the recently proposed general method of the tensor renormalization
group method\cite{TRG}, we express the partition function in terms of a product of 
vertices (sometimes called tensors, the origin of the name, tensor renormalization 
group).
\Be
Z = \sum_{abcdefg\cdots}T_{abcd}T_{edfg}\cdots
\Ee
The vertex $T$ is defined on each dual site and has 4 indices representing
domain wall state on 4 dual links. We assign 0 for no-domain wall and 1 for
existence of domain wall. 
Then the components of $T$ are given by
\Ba
T_{0000} &=& \exp(2\beta J) \ ,\nonumber\\
T_{0101} &=& T_{0110} =1\ ,\\
T_{1111} &=& \exp(-2\beta J) \ ,\nonumber 
\Ea
where the index cyclic symmetry is assumed.

The coarse graining procedure consists of two steps. The first step is
to break $T$-vertex into a product of two $S$-vertex.
From the Feynman diagram view, 
this step can be seen as introducing an intermediating boson for the 4-fermi
weak interactions.
\Be
\vcenter{\hbox{\includegraphics[bb=0 0 160 170, width=20mm]{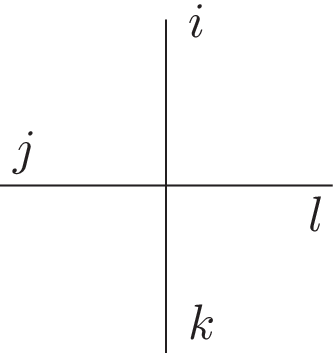}}}
\hskip5mm
\simeq
\hskip5mm
\vcenter{\hbox{\includegraphics[bb=0 0 167 180, width=20mm]{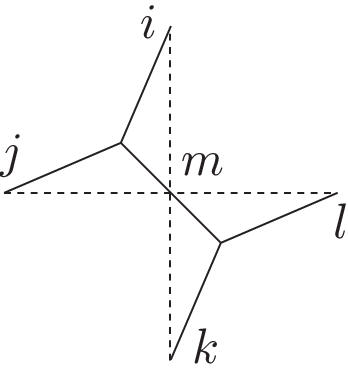}}}
\Ee
The intermediating line will become a coarse grained dual link and 
coarse grained domain wall states are defined on it.
If we take 4 states for that, we do not lose any information. 
However we like to set up a finite dimensional system of the renormalization group
transformation and we have to discard some degrees of freedom here.
Then we take only 2 states on the intermediate states.
It means this decomposition of $T$-vertex loses information, and this is the
approximation of this type of renormalization group transformation.

The 2-state approximation must be the lowest approximation and we will 
analyze its RGT in detail.
Which two states among totally $2\times 2=4$ states should be picked up ?
There are two ways to determine it. One is a physical condition that
coarse grained states should obey notions of the coarse grained domain 
walls. The other is a practical condition that we should minimize the
information loss of this procedure.
Fortunately, we prove below that these two conditions are consistent
for some region of parameters which we call physical region, and it is respected
by the physical finite temperature system.

\begin{figure}
\begin{center}
\includegraphics[bb=0 0 385 116, width=11cm]{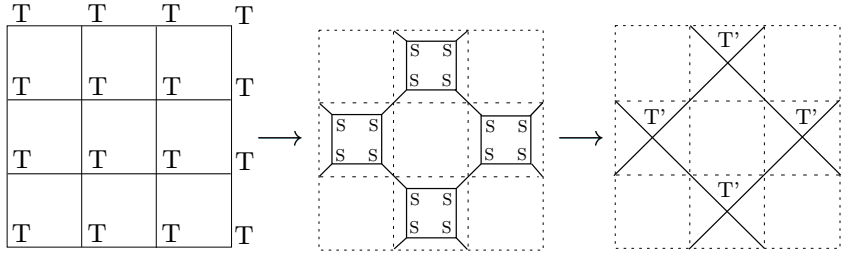}
\nobreak
\caption{Integrate out $S$-vertex to define renormalized $T$-vertex}
\label{fig:rgttwo}
\end{center}
\end{figure}

The second step of RGT is to integrate out all $S$-vertices and will make the
system described only by new variables, the intermediate states.
The new variables are nothing but the coarse grained domain walls on the
newly made coarse grained dual links (Fig.\ref{fig:rgttwo}).
These two steps define RGT with scale factor of $\sqrt{2}$.

\section{Analytic expression of RGT}

We express $T$-vertex (with four legs) by a $4\times4$ matrix.
We combine two legs each into row or column index of the matrix.
The correspondence rules are
\Be
\{00\} \Rightarrow \{1\}\ ,\ 
\{11\} \Rightarrow \{2\}\ ,\ 
\{01\} \Rightarrow \{3\}\ ,\ 
\{10\} \Rightarrow \{4\}\ ,\ 
\Ee
Then we assume the texture of $T$ matrix as follows:
\Be
{T}=\bordermatrix{
& 00 & 11 & 01 & 10 \cr
00 & 1 & b & 0 & 0  \cr 
11 & b & c & 0 & 0  \cr 
01 & 0 & 0 & a & b  \cr
10 & 0 & 0 & b & a  \cr  
 }\ .\label{eq:texture}
\Ee
where a,b,c are real positive parameters obeying
the conditions defined by
\Ba
0 &<& a \leq b <1\ ,\nonumber\\
0 &<&c <b \ ,\label{eq:physcon}\\
2b &<& c+1\ ,\nonumber
\Ea
which we call the physical region condition.
In general, renormalization group transformation modifies $T_{11}$ component and 
we always normalize the matrix by dividing all elements by a constant
so that $T_{11}$ should be 1.

These two properties, the texture and physical region condition, are
satisfied by the initial $T$ matrix given by
\Ba
a&=&b=\alpha\ ,\nonumber\\
c&=&\alpha^2\ ,\\
\alpha &=& \exp(-2\beta J)< 1 .\nonumber
\Ea
Now will prove that our RGT does not break these two properties and therefore
all flows starting from the physical region have the same texture and belongs
to the physical region.

First we calculate eigenvalues of $T$. 
\Ba
\lambda_{1,2} &=&\frac{1}{2}\left( 1+c \pm \sqrt{(1-c)^2 +4b^2} \right) \ ,\nonumber\\
\lambda_{3,4} &=&a\pm b \ .
\Ea
We examine the absolute values of these 4 eigenvalues and select 
two larger components according to the policy of TRG.
Using the physical region condition, we can prove the following inequalities
\Ba
\lambda_1 &>& b\ ,\ \ 
\lambda_3 > b\ ,\nonumber\\
|\lambda_4| &<& b\ ,\ \ 
|\lambda_2|<b\ .
\Ea
Thus we take two eigenvalues $\lambda_1,\lambda_3$ to define $S$ vertex.
Note that these two eigenvalues are positive.

Now we can split $T$-vertex into a product of two $S$-vertices.
First we introduce two orthogonal $2\times2$ matrices $R_{1,2}$ which 
diagonalize $T$ as follows:
\Be
T= 
\left( 
\begin{array}{cccc}
 1 & b & 0 & 0  \\
 b & c & 0 & 0  \\ 
 0 & 0 & a & b  \\
 0 & 0 & b & a  \\  
\end{array} 
\right)
=
\left( 
\begin{array}{cc}
{\rm R_1} & 0 \\ 
0 & {\rm R_2}\\   
\end{array} 
\right)
\left( 
\begin{array}{cccc}
\lambda_1 & 0 &0 & 0 \\
0 & \lambda_2 & 0& 0\\
0 & 0 & \lambda_3 & 0 \\
0 & 0 &0 & \lambda_4
\end{array}
\right)
\left( 
\begin{array}{cc}
{\rm R_1^t} & 0 \\ 
0 & {\rm R_2^t}\\   
\end{array}
\right)
 \ ,
\Ee
where $R_{1,2}$ are defined by angles $\theta_{1,2}$
\Ba
{\rm R_1}&=&
\left(
\begin{array}{cc}
\cos \theta_1 & -\sin \theta_1 \\ 
\sin \theta_1 & \cos \theta_1 \\   
 \end{array}\right)\ ,\nonumber\\
{\rm R_2}&=&\left( \begin{array}{cc}
\cos \theta_2 & -\sin \theta_2 \\ 
\sin \theta_2 & \cos \theta_2 \\   
 \end{array}\right)\ .
\Ea
Therefore we have the following decomposition:
\Be
T=S^{\mbox{\tiny A}} \cdot S^{\mbox{\tiny B}}\ ,
\Ee
where $S^{\mbox{\tiny A}}$ is $4\times 2$ matrix and 
$S^{\mbox{\tiny B}}$ is $2\times 4$ matrix given by
\Ba
S^A&=&\left( \begin{array}{cc}
\sqrt{\lambda_1}c_1 & 0 \\
\sqrt{\lambda_1}s_1 & 0\\
 0 & \sqrt{\lambda_3}c_2 \\
 0 & \sqrt{\lambda_3}s_2 
\end{array}\right)\ , \nonumber\\
S^B&=& \left( \begin{array}{cccc}
\sqrt{\lambda_1}c_1 & \sqrt{\lambda_1}s_1 & 0 & 0 \\
0 & 0 & \sqrt{\lambda_3}c_2 & \sqrt{\lambda_3}s_2 \\
\end{array}\right)\ .
\Ea
It should be noted here that these $S$-vertices satisfy
$S^{\mbox{\tiny A}}=(S^{\mbox{\tiny B}})^t$ 
and they are essentially the same.
This decomposition is nothing but the singular value decomposition 
which is frequently used in order to pick up important degrees of freedom.
Hereafter we use simple notations as follows:
\Be
s_1 = \sin \theta_1\ ,\ \ c_1 = \cos \theta_1\ ,\ \ 
s_2 = \sin \theta_2\ ,\ \ c_2 = \cos \theta_2\ .
\Ee

\section{Feynman rules to calculate RGT}

The $T$-vertex decomposition is expressed as
\Be
\vcenter{\hbox{\includegraphics[bb=0 0 146 142, height=15mm]{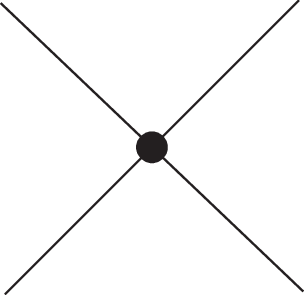}}\vskip-1mm}
\ =\
\vcenter{\hbox{\includegraphics[bb=0 0 246 143, height=15mm]{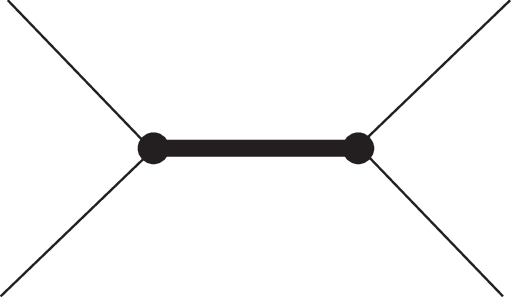}}\vskip-1mm} 
\Ee
\vskip-4mm\nobreak
\centerline{\hbox to 20mm{\hfil $T$\hfil} \hspace{9mm}
\hbox to 14mm{$S^{\mbox{\tiny A}}$\hfil $S^{\mbox{\tiny B}}$}\hskip8mm} 
\noindent
where the inner thick line represents two possible states.
We assign names to these two states to be 0 ($\lambda_1$ component) and
1 ($\lambda_3$ component) respectively.
This assignment is essential for interpreting the coarse grained 
domain wall variables, which will be clear soon.
The $S$-vertex gives the following Feynman rules:

\vbox{
\Be
\vcenter{\hbox to 120mm{\includegraphics[bb=0 0 172 186, width=27mm]{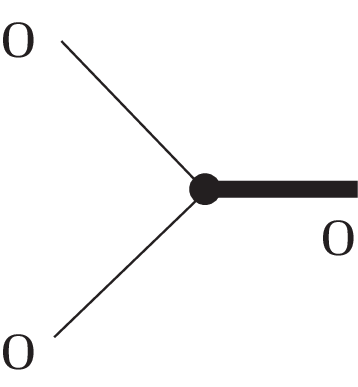}\hfil
\includegraphics[bb=0 0 173 190, width=27mm]{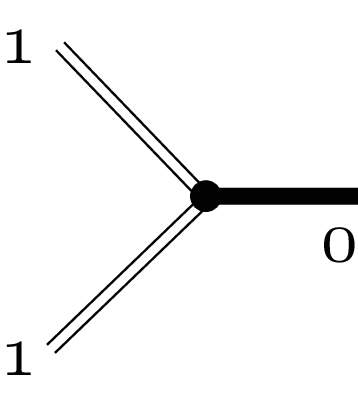}\hfil
\includegraphics[bb=0 0 173 186, width=27mm]{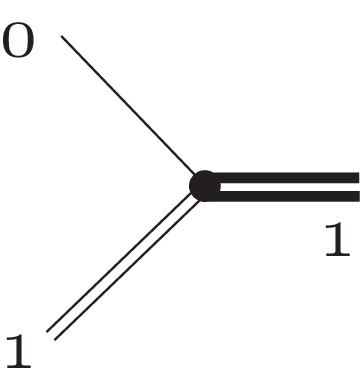}\hfil
\includegraphics[bb=0 0 174 186, width=27mm]{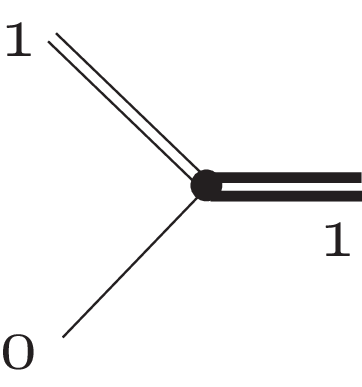}}}
\Ee
\nobreak
\vskip-5mm
\nobreak
\centerline{\hbox to 108mm{$\sqrt{\lambda_1}c_1$ 
\hfil $\sqrt{\lambda_1}s_1$ \hfil $\sqrt{\lambda_3}c_2$ \hfil $\sqrt{\lambda_3}s_2$}}
}

\vspace{3mm}

\noindent
where we draw single line for 0-state and double line for 1-state.
These rule applies to all legs.
Now we see $S$-vertex conserves the number of legs of double line (1-state), thus 
conserves the domain wall.

The renormalization group transformation is defined by one-loop 
diagrams and they are calculated as follows:
\Ba
\vcenter{\hbox{\includegraphics[bb=0 0 146 146, width=25mm]{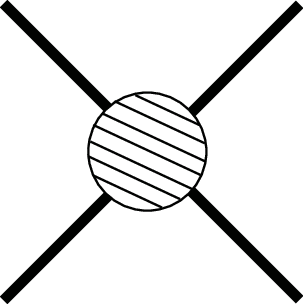}}\vskip-1mm}\ 
\ &=&\ 
\vcenter{\hbox{\includegraphics[bb=0 0 219 215, width=25mm]{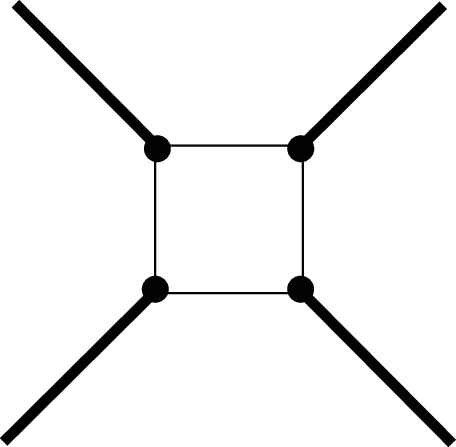}}\vskip-1mm}\ 
\ +\ 
\vcenter{\hbox{\includegraphics[bb=0 0 218 215, width=25mm]{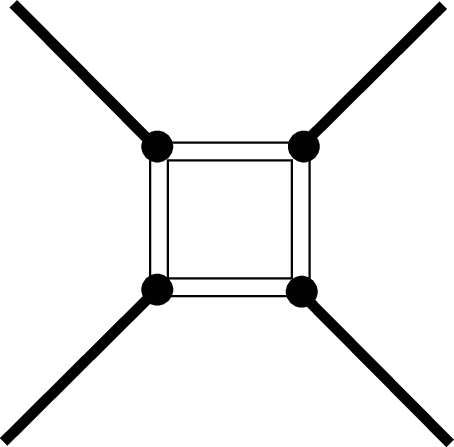}}\vskip-1mm}\ 
\ =\ \lambda_1^2 \left( c_1^4 +s_1^4 \right)\ ,\\ \nonumber \\
\vcenter{\hbox{\includegraphics[bb=0 0 153 151, width=25mm]{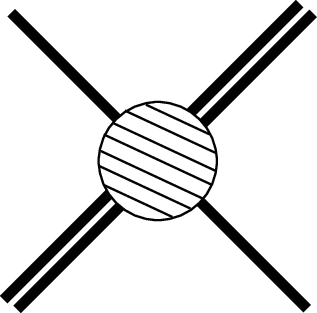}}\vskip-1mm}\ 
\ &=&\ 
\vcenter{\hbox{\includegraphics[bb=0 0 221 219, width=25mm]{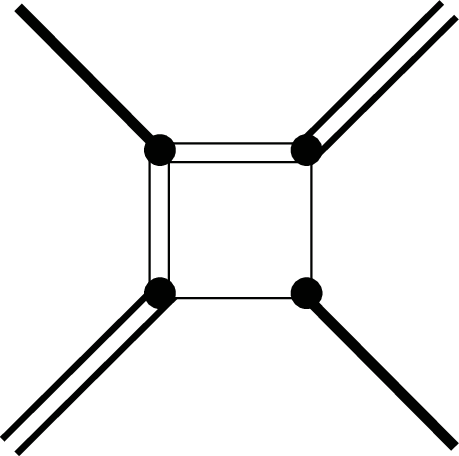}}\vskip-1mm}\ 
\ +\ 
\vcenter{\hbox{\includegraphics[bb=0 0 221 220, width=25mm]{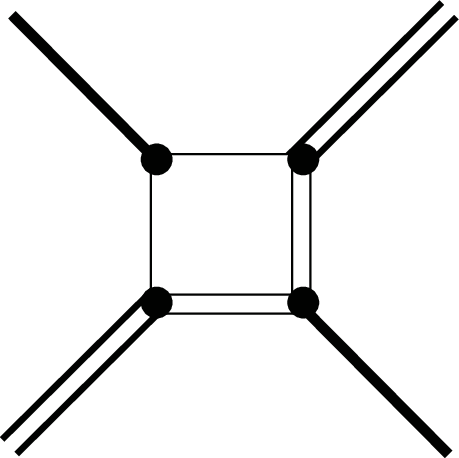}}\vskip-1mm}\ 
\ =\ 2\lambda_1 \lambda_3 c_1 s_1 c_2 s_2\ ,\\ \nonumber \\
\vcenter{\hbox{\includegraphics[bb=0 0 147 155, width=25mm]{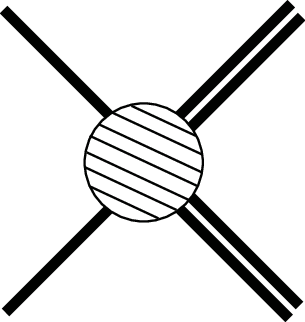}}\vskip-1mm}\ 
\ &=&\ 
\vcenter{\hbox{\includegraphics[bb=0 0 219 222, width=25mm]{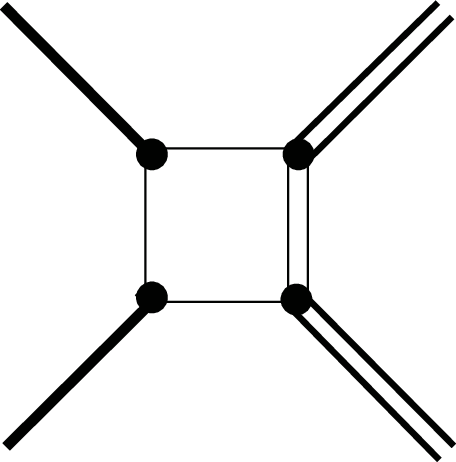}}\vskip-1mm}\ 
\ +\ 
\vcenter{\hbox{\includegraphics[bb=0 0 220 224, width=25mm]{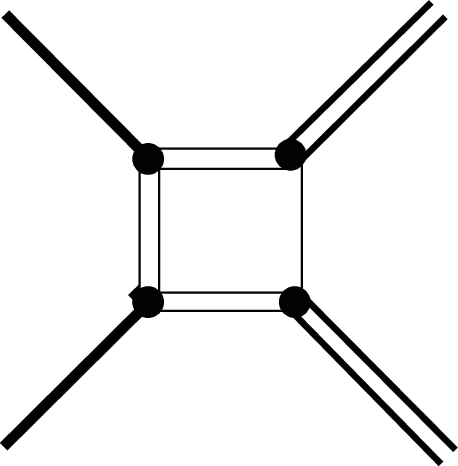}}\vskip-1mm}\ 
\ =\ \lambda_1 \lambda_3 c_2 s_2 \left( c_1^2 +s_1^2 \right)\ ,\\ \nonumber \\
\vcenter{\hbox{\includegraphics[bb=0 0 153 152, width=25mm]{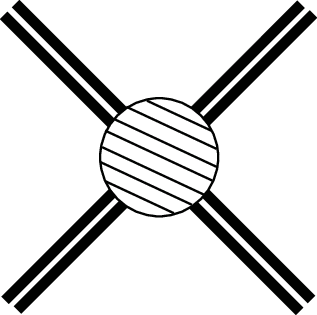}}\vskip-1mm}\ 
\ &=&\ 
\vcenter{\hbox{\includegraphics[bb=0 0 221 222, width=25mm]{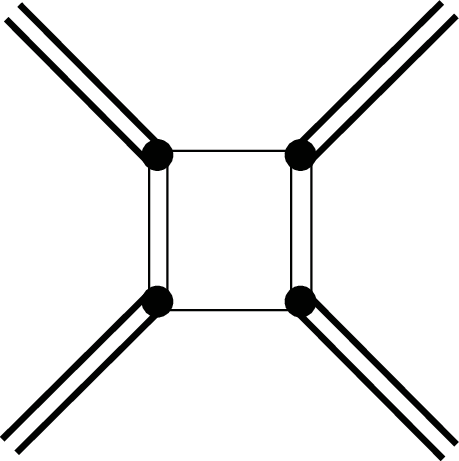}}\vskip-1mm}\ 
\ +\ 
\vcenter{\hbox{\includegraphics[bb=0 0 224 221, width=25mm]{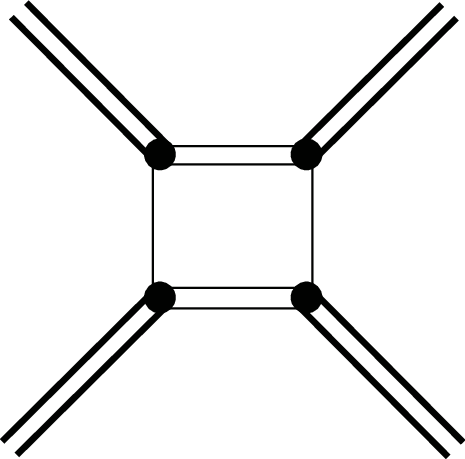}}\vskip-1mm}\ 
\ =\ 2\lambda_3^2 c_2^2 +s_2^2 \ .
\Ea
Note that the conservation of domain wall prohibits other diagrams.
These amplitudes give the renormalized $T$-vertex.
Taking account of the discrete rotational symmetry,
\Be
\vcenter{\hbox{\includegraphics[bb=0 0 147 155, width=18mm]{figs/F3.eps}}\vskip-1mm}
=
\vcenter{\hbox{\includegraphics[bb=0 0 155 147, width=18mm]{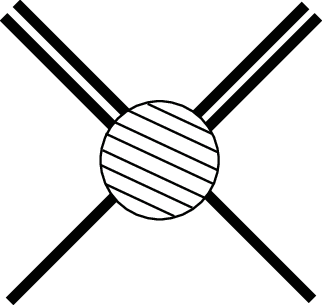}}\vskip-1mm}
=
\vcenter{\hbox{\includegraphics[bb=0 0 146 153, width=18mm]{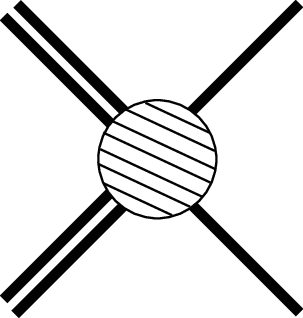}}\vskip-1mm}
=
\vcenter{\hbox{\includegraphics[bb=0 0 153 151, width=18mm]{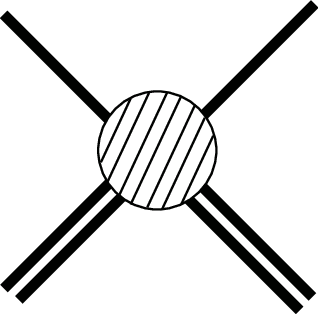}}\vskip-1mm}
\ ,\ 
\vcenter{\hbox{\includegraphics[bb=0 0 153 151, width=18mm]{figs/F2.eps}}\vskip-1mm}
=
\vcenter{\hbox{\includegraphics[bb=0 0 152 152, width=18mm]{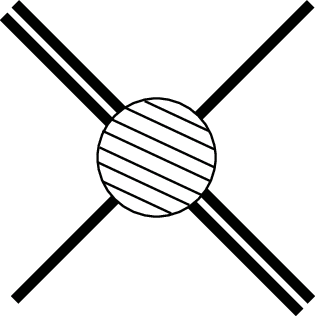}}\vskip-1mm}\ ,
\Ee
we have the renormalized $T$-vertex ($T^\prime$) 
\Be
T^{\prime}=\left(\begin{array}{cccc}
1       & b^\prime &0         & 0        \\
b^\prime& c^\prime & 0        & 0        \\
0       & 0        & a^\prime & b^\prime \\
0       & 0        & b^\prime & a^\prime
\end{array}\right)
\ ,
\Ee
where we made additional total renormalization of the matrix elements so that
$T^\prime_{11}$ should be equal to 1.

Thus we have proved that RGT conserves the texture defined in Eq.(\ref{eq:texture}).
This comes from the property that the $S$-vertex conserves the domain wall with
our assignment of coarse grained domain walls. 
Also the discrete rotational symmetry is necessary.
In fact the physical region condition (\ref{eq:physcon}) is not necessary for this 
texture conservation. Strictly speaking the necessary and sufficient condition is 
that each of the two coarse grained components should come from the upper block 
and the lower block respectively. The physical region condition corresponds to 
another property
that each of the two largest values among 4 eigenvalues are in the upper block 
and the lower block respectively. 
Thus the physical region condition assures that both of the best 
approximate coarse graining and the domain wall conservation holds simultaneously.

We should discuss here the notion of coarse grained domain walls.
The $S$-vertex defines the relation of the micro domain walls
and macro (coarse grained) domain wall. 
The micro domain wall does not always generates macro domain wall.
The small loop of the micro domain wall does not correspond to the macro domain wall, 
whereas the domain walls connecting two coarse grained vertices
(domain walls residing on the coarse grained links) are defined as
macro domain walls.
These interpretation naturally defines the coarse graining of the domain 
walls and renormalization of the $T$-vertex.

\section{Renormalization Transformation}

The renormalization transformation for parameters $a,b,c$ are 
written down as follows:
\Ba
a^\prime &=& \frac{2\lambda_3}{\lambda_1} \frac{c_1 s_1 c_2 s_2}{ c_1^4  +s_1^4}\ ,\ 
\nonumber \\
b^\prime &=& \frac{\lambda_3}{\lambda_1} \frac{c_2 s_2}{ c_1^4  +s_1^4}\ ,\ 
\\
c^\prime &=& \frac{2\lambda_3^2}{\lambda_1} \frac{c_2^2 s_2^2}{ c_1^4  +s_1^4}\ .
\nonumber
\Ea
Angles are obtained as a function of parameters $a,b,c$
\Ba
\sin 2\theta_1 &=& \frac{2b}{\lambda_1 -\lambda_2}=\frac{2b}{ \sqrt{(1-c)^2 + 4b^2}}\ ,
\nonumber\\
\sin 2\theta_2&=&1\ .
\Ea
Using these angles we rewrite the transformation
\Ba
a^\prime &=& \left( \frac{\lambda_3}{\lambda_1} \right)\ \frac{s}{2-s^2}\ ,\ 
\nonumber \\
b^\prime &=& \left( \frac{\lambda_3}{\lambda_1} \right)\ \frac{1}{2-s^2}\ ,\ 
\\
c^\prime &=& \left( \frac{\lambda_3}{\lambda_1} \right)^2 \frac{1}{2-s^2}\ ,
\nonumber
\Ea
where eigenvalues and diagonalization angles are
\Ba
\lambda_1 &=& \frac{1}{2} \left( 1+c+\sqrt{(1-c)^2 +4b^2} \right)\ ,\nonumber\\
\lambda_3 &=& a+b\ , 
s = \frac{2b}{\sqrt{(1-c)^2 +4b^2}} \ .\label{eq:coparameters}
\Ea
This completes the explicit analytic formula of the domain wall renormalization group
transformation.

Finally we check the physical region condition for renormalized parameters.
Due to the inequality $0 <a \leq b$, we have
\Be
\lambda_3 < 2b\ .
\Ee 
On the other hand, we define a function $f(\lambda)$
\Be
f(\lambda)=\lambda^2 -(1+c)\lambda +c -b^2=0
\Ee
where $\lambda_1$ is a larger root of this equation.
We have the following inequality 
\Be
f(2b) = (1-b)(c-b)+b(2b-c-1) < 0
\Ee
and it proves $\lambda_1 > 2b$. Then we have 
\Be
\frac{\lambda_3}{\lambda_1} < 1\ .
\Ee
Also using $0 \leq s \leq 1$, 
we can finally prove 
\Ba
0 &<& a^\prime \leq b^\prime <1 \ ,\nonumber\\
0 &<& c^\prime < b \ ,\\
2b^\prime &<& c^\prime + 1\ .\nonumber
\Ea
Therefore we have proved that the physical region condition (\ref{eq:physcon})
is conserved by the renormalization transformation.
Then the above analytic forms of the transformation and physical region
condition always hold for any flows starting from a point in the physical region.

\section{Fixed points}

According to the standard procedure, we look for fixed points of the 
renormalization transformation.
We can see there are two trivial fixed points.

\ \hbox to 52mm{Low temperature fixed point:\hfil}$\{a=0\ ,\ b=0\ ,\ c=0\}$

\ \hbox to 52mm{High temperature fixed point:\hfil}$\{a=1\ ,\ b=1\ ,\ c=1\}$

\noindent
These fixed points are not in the physical region but on the boundary of the
region. They are both infrared fixed points which almost all flows
(except for those on the critical surface) in the physical
region approaches in the infrared limit.
We call them low or high temperature fixed point respectively because
they are equal to the physical initial $T$-vertex with $\alpha=0
 (\beta\rightarrow\infty)$ or $\alpha=1 (\beta\rightarrow 0)$.

We check the eigenvalues of the renormalization transformation 
in the neighborhood of these trivial fixed points and find
that all eigen values are less than unity for both fixed points.
Thus, we have two infrared fixed points at the boarder of the physical 
region. Then there must necessarily exist a critical surface 
which divides the space into two subspace (different phases). 
Phases are characterized by these infrared fixed point.
On the critical surface there must exist at least one 
(non-trivial) fixed point.

The non-trivial fixed point satisfies the following set of equations
\Ba
a &=& \left( \frac{\lambda_3}{\lambda_1}\right) \ \frac{s}{2-s^2} \ ,\nonumber\\
b &=& \left( \frac{\lambda_3}{\lambda_1}\right) \ \frac{1}{2-s^2} \ ,\label{eq:fixed}\\
c &=& \left( \frac{\lambda_3}{\lambda_1}\right)^2 \frac{1}{2-s^2} \ , \nonumber
\Ea
where $\lambda_1, \lambda_3, s$ are defined in Eq.(\ref{eq:coparameters}).
Noticing that parameters $s$ and $\lambda_1$ 
are functions of $b, c$ only, we first solve $a$D
Dividing the first equation by the second equation in the above, we have
\Be
a= bs =  \frac{2b^2}{\sqrt{(1-c)^2+4b^2}}\ . \label{eq:a8}
\Ee
Also dividing the second equation by the third equation in Eq.(\ref{eq:fixed}),
we have
\Be
b^2(2-s^2)=c\ . \label{eq:bc2}
\Ee
and we solve it with respect to $b$,
\Be
b=\frac{1}{2} \sqrt{2c-(1-c)^2 +\sqrt{4c^2 +(1-c)^4}}\ . \label{eq:b9}
\Ee
Next adding the first and second equations in 
Eq.(\ref{eq:fixed}) we have
\Be
\lambda_1 (2-s^2) = 1+s\ .\label{eq:bc1}
\Ee
Substituting Eq.(\ref{eq:b9}) into the above equation, we have an equation for $c$.
\Be
136c^5 -145c^4 +116c^3- 54c^2 +12c -1=0\ .
\Ee
This 5th order polynomial equation is easily found to have only one real root.
We finally get the unique non-trivial fixed point
\Ba
c^*&=&0.238902743\ ,\nonumber\\
b^*&=&0.402938077\ ,\\
a^*&=&0.292942734\ .\nonumber
\Ea

\begin{figure}
\begin{center}
\includegraphics[bb=0 0 318 222, width=14cm]{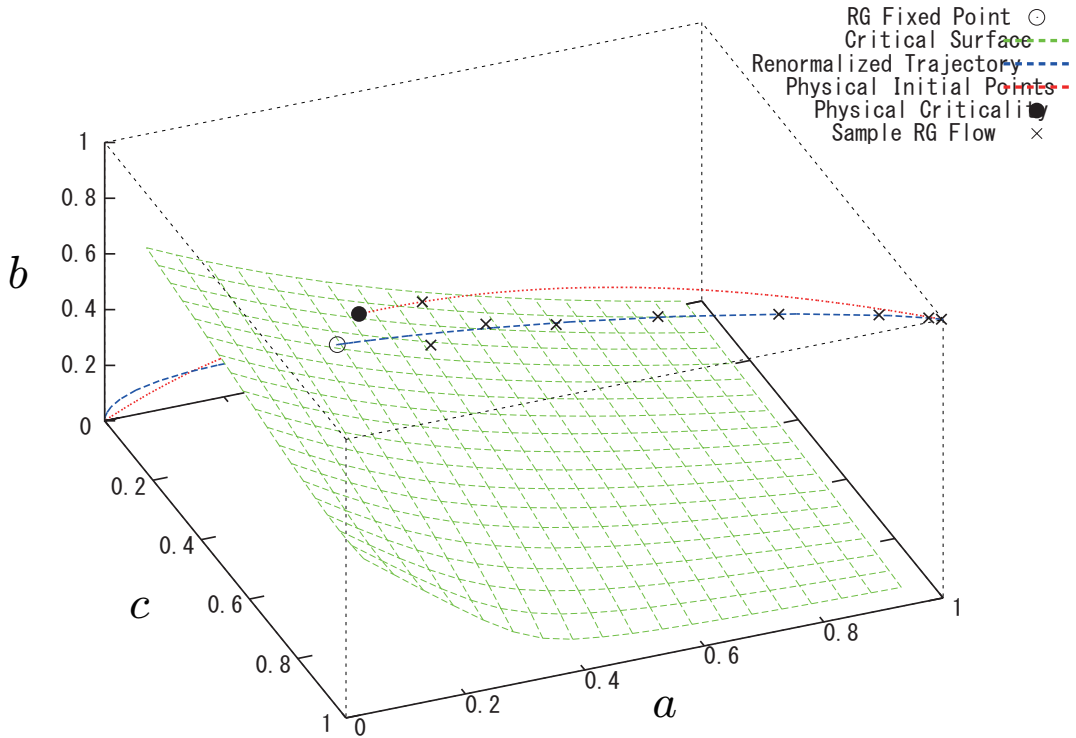}
\nobreak
\caption{Phase structure of domain wall RGT}
\label{fig:d4c}
\end{center}
\end{figure}

Numerical results for the critical surface and the renormalized 
trajectory are drawn in Fig.\ref{fig:d4c}.
Note that we plotted here whole region of $0<a,b,c<1$.
There is no singular behavior at the boundary of the physical 
region condition once the renormalization transformation is given.
Also the physical initial points and a sample renormalization group
flow are plotted there.
The cross point of physical initial points and the critical 
surface determines the critical temperature.
We have the criticality in terms of $\alpha$
\Be
\alpha_{\mbox{\tiny c}}=0.37036\ ,\ \ 
\Ee
which should be compared with the exact value 
$\alpha_{\mbox{\tiny c}}(\mbox{Exact})=\sqrt{2}-1=0.41421$.
Our result suffers about 10\% error.

We linearize the renormalization transformation around the non-trivial
fixed point and find the eigenvalues
\Be
\{1.4224236, -0.29079698, 0\}\ .
\Ee
There is only one eigenvalue $\lambda=1.4224236$ which is larger than unity.
Thus the fixed point has only one relevant operator as is reduced by the general argument.
According to the standard method\cite{Wilson} to evaluate the critical 
exponent $\nu$ of the correlation length divergence, we have
\Be
\nu = \frac{\log \sqrt{2} }{\log \lambda} = 0.98357\ .
\Ee
Here we have used that the renormalization transformation changes the lattice 
scale by factor $\sqrt{2}$.
This result seems extremely good, less than 2\% error compared to the exact value 
$\nu = 1$, considering that our domain wall renormalization group has only 
two states on the coarse grained link and only 3-dimensional interaction space
has been used. 

\begin{figure}
\hbox to \hsize{\hfil\includegraphics[bb=0 0 343 245, width=12cm]{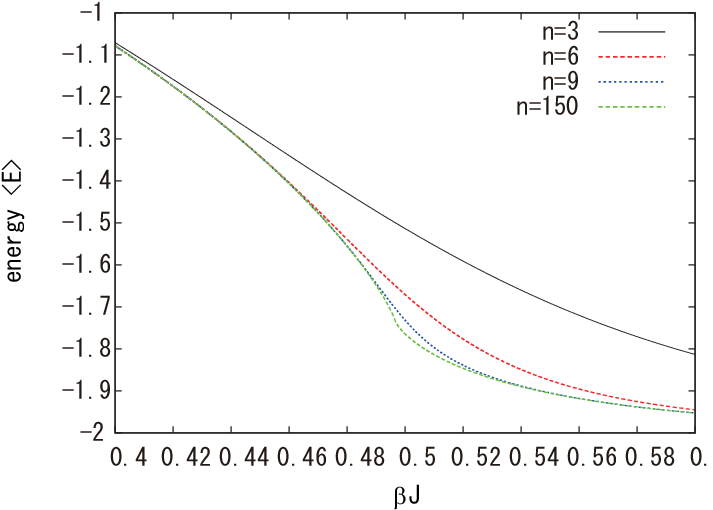}\hfil}
\vspace{-3mm}\caption{Mean energy as a function of $\beta J$}
\label{fig:meanenergy}

\vspace{15mm}

\hbox to \hsize{\hfil\includegraphics[bb=50 50 410 292, width=12cm]{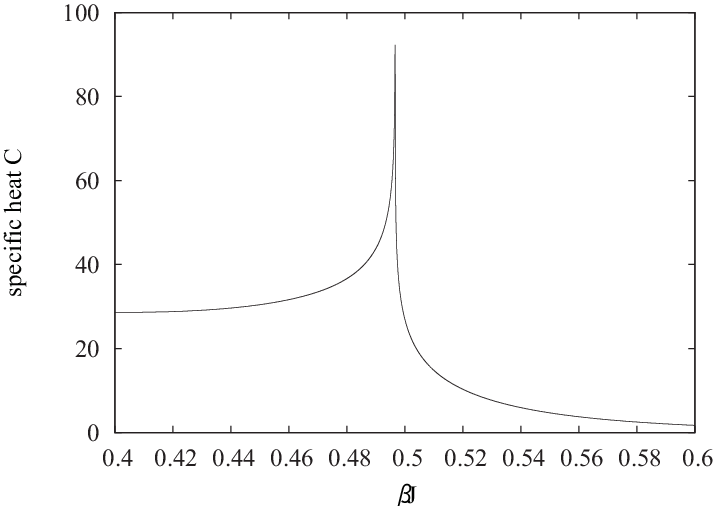}\hfil}
\vspace{-3mm}\caption{Specific heat as a function of $\beta J$}
\label{fig:specificheat}
\end{figure}

We proceed to investigate the behavior of the mean energy and the 
specific heat.
The partition function is calculated by tracing out the $T$-vertex.
Note that we have to take account of the additional total normalization
factor which is introduced at each renormalization step
to make $T_{11}$ to be unity.
We denote the factor by $C_{n}$ which is 
used to divide $T$ at the $n$-th renormalization step.
After $n$-times renormalization transformation, the $T$-vertex is 
renormalized to become $T^{(n)}$ and we define
\Be
Z^{(n)} (\beta) = C_{\mbox{\tiny tot}} \ T^{(n)}_{abab}
= C_0^{2^n} C_1^{2^{n-1}} C_2^{2^{n-2}} \cdots C_n \ T^{(n)}_{abab}\ ,
\Ee
which is the partition function of the system with size
$2^n$ (the total number of sites).

We calculate the mean energy and the specific heat per site.
The results are shown in Figs.\ref{fig:meanenergy} and \ref{fig:specificheat}.
In the mean energy plots, increasing $n$, there start to appear 
a would-be singular point near $\beta J \simeq 0.5$. 
The specific heat would diverge at the point due to non-analyticity 
at $n\rightarrow\infty$.
The divergence behavior of the specific heat 
is almost completely logarithmic as is derived by the scaling relation.

We can increase the number of states defined on the coarse grained link.
As a next approximation we set up 4-state version of the domain wall 
renormalization group.
The corresponding $T$-vertex has $4^4=256$ components.
The texture analysis shows that only 25 components are non-vanishing in the 
procedure of the renormalization group transformation.
The details of our study will be reported 
in a full paper\cite{Aoki-Kobayashi-Tomita09b}.
Here we only mention some results.
The critical temperature is obtained as
\Be
\alpha_c[\mbox{4-state}] = 0.42205\ ,
\Ee
and the error is now less than 2\%. This is a great improvement 
compared to the 2-state result.
As for the correlation length critical exponent, we have
\Be
\nu[\mbox{4-state}] = 0.98359\ ,
\Ee
and this is almost equal to the 2-state result.

In the 4-state version RGT, there are some 
subtle issues to be discussed in detail.
For example, notion of the coarse grained domain wall is not
trivial like 2-state version. 
To evaluate the magnetic quantities like the magnetization and
the susceptibility, we need to introduce external source field.
This will break the symmetry property we have used in this article
and it needs additional consideration.

We would like to thank fruitful discussion with H.~Suzuki who 
was collaborating with us in the initial stage of this work.
This research was partially supported by the Ministry of Education, 
Science, Sports and Culture, Grant-in-Aid for 
Scientific Research (B),17340070,2007.


\begin{thebibliography}{0}

\bibitem{Wilson}
K.~G.~Wilson,
Rev. Mod. Phys. {\bf 47}  (1975) 773.

\bibitem{TRG}
M.~Levin and C.~P.~Nave,
Phys. Rev. Lett. {\bf 99} (2007) 120601.

\bibitem{TRG-triangle}
M.~Hinczewski and N.~Berker,
Phys. Rev. E {\bf 77} (2008) 011104.

\bibitem{Tensor-entanglement-square-transverse-ising}
Z.-C.~Gu, M.~Levin and X.G. Wen,
Phys. Rev. B {\bf 78} (2008) 205116 and 
arXiv:0806.3509 [cond-mat.str-el].

\bibitem{Aoki-Kobayashi-Tomita09b}
K-I.~Aoki, T.~Kobayashi and H.~Tomita in preparation.
\end{thebibliography}
\end{document}